\documentclass{PoS}
\usepackage{amsmath}

\title{Minimal doubling and point splitting}

\ShortTitle{Minimal doubling and point splitting}

\author{\speaker{Michael Creutz}%
        \thanks{I am grateful to the Alexander von Humboldt Foundation
          for support for my ongoing visits to the University of
          Mainz.  This manuscript has been authored under contract
          number DE-AC02-98CH10886 with the U.S.~Department of Energy.
          Accordingly, the U.S. Government retains a non-exclusive,
          royalty-free license to publish or reproduce the published
          form of this contribution, or allow others to do so, for
          U.S.~Government purposes.}\\
        Brookhaven Lab\\
        E-mail: \email{mike@latticeguy.net}}

\abstract{
Minimally-doubled chiral fermions have the unusual property of a
single local field creating two fermionic species.  Spreading the
field over hypercubes allows construction of combinations that
isolate specific modes.  Combining these fields into bilinears
produces meson fields of specific quantum numbers.
}

\FullConference{The XXVIII International Symposium on Lattice Filed Theory\\
        June 14-19,2010\\
        Villasimius, Sardinia Italy}

\begin{document}

\section{Introduction}

The classic Nielsen-Ninomiya theorem \cite{Nielsen:1980rz} states that
for a local lattice action to have an exact chiral symmetry, it must
describe an even number of flavors.  Within the fundamental Brillouin
zone, the fermion propagator should possess an even number of poles.
The argument is topological, involving a non-trivial mapping
associated with each pole, and this mapping must unwrap elsewhere in
the zone. For two space-time dimensions this is a wrapping of a
complex number around a circle, while for four dimensions the chiral
Dirac operator involves a mapping of quaternions around a sphere $S_3$
\cite{Creutz:2007af}.

Several chiral lattice actions satisfying the minimal condition of
$N_f=2$ flavors are known.  Some time ago Karsten
\cite{Karsten:1981gd} presented a simple form by inserting a factor of
$i\gamma_4$ into a Wilson like term for space-like hoppings.  A slight
variation appeared in a discussion by Wilczek \cite{Wilczek:1987kw} a
few years later.  More recently, I developed a four dimensional action
motivated by the analogy with two dimensional graphene.  Since then a
few more variations were presented by Borici \cite{Borici:2007kz},
Bedaque et al. \cite{Bedaque:2008jm}, and Kimura and Misumi
\cite{Kimura:2009di}.  An effort to classify these various approaches
appears in \cite{Creutz:2010cz}.

The main potential advantage of these approaches lies in their
ultra-locality.  They should be extremely fast in simulations while
protecting masses from additive renormalization and helping control
mixing of operators with different chirality.  The approach also
avoids the uncontrolled errors associated with the rooting
approximation \cite{rooting}.  On the other hand, all
minimally-doubled actions presented so far have the disadvantage of
breaking the usual lattice hyper-cubic symmetry.  With interactions,
this will lead to the necessity of renormalization counter-terms that
also violate this symmetry.  The extent to which this complicates
simulations remains to be investigated.

Here I discuss a point-splitting method for separating the effects of
the two flavors which can be created by a single fermion field.  For
this I will concentrate on the Karsten/Wilczek form as somewhat
simpler, but the method should be easily extended to other
minimally-doubled formulations.

In the next section I review this action and discuss some of its
properties.  In section \ref{perturbative} I discuss recent
developments on perturbative renormalization and the concomitant
counter-terms. Section \ref{splitting} introduces the proposed point
splitting and discusses some of the combinations that create physical
particles.  Section \ref{effective} uses a simple effective Lagrangian
argument to discuss some of the lattice artifacts that may occur.
Finally section \ref{conclusions} summarizes the basic conclusions.

\section{Karsten/Wilczek fermions}
\label{kw}
I concentrate on a minimally-doubled fermion action which is a slight
generalization of those presented by Karsten \cite{Karsten:1981gd} and
Wilczek \cite{Wilczek:1987kw}.  The fermion term in the lattice action
takes the form $\overline\psi D\psi$.  For free fermions I start with
the momentum space form
\begin{equation}
\label{pspace}
D(p)=
i\sum_{i=1}^3 \gamma_i \sin(p_i)
+{i\gamma_4\over \sin(\alpha)}\left(
\cos(\alpha)+3
-\sum_{\mu=1}^4\cos(p_\mu)
\right)
\end{equation}
As a function of the momentum $p_\mu$, the propagator
$D^{-1}(p)$ has two poles, located at $\vec p = 0$, $p_4=\pm \alpha$.
Relative to the naive fermion action, the other doublers have been
given a large ``imaginary chemical potential.''  The parameter
$\alpha$ allows adjusting the relative positions of the poles.  The
original Karsten/Wilczek actions correspond to $\alpha=\pi/2$.

This action maintains one exact chiral symmetry, manifested in the
anti-commutation relation $[D,\gamma_5]_+=0$.  The two species,
however, are not equivalent, but have opposite chirality.  To see
this, expand the propagator around the two poles and observe that one
species, that corresponding to $p_4=+\alpha$, uses the usual gamma
matrices, but the second pole gives a proper Dirac behavior using
another set of matrices $\gamma_\mu^\prime=\Gamma^{-1}\gamma_\mu
\Gamma$.  The Karsten/Wilczek formulation uses
$\Gamma=i\gamma_4\gamma_5$, although other minimally-doubled actions
may involve a different transformation.  After this transformation
$\gamma_5^\prime=-\gamma_5$, showing that the species rotate
oppositely under the exact chiral symmetry and thus this symmetry is
``flavored.''  One can think of the physical chiral symmetry as that
generated in the continuum theory by $\tau_3\gamma_5$.

It is straightforward to transform the momentum space action in
Eq.~(\ref{pspace}) to position space and insert gauge fields
$U_{ij}=U_{ji}^\dagger$ on the links connecting lattice sites.
Explicitly indicating the site indices, the Dirac operator becomes
\begin{equation}
 D_{ij}=
U_{ij}\sum_{\mu=1}^3 \gamma_i{\delta_{i,j+e_\mu}
-\delta_{i,j-e_\mu}\over
2}
+
{i\gamma_4\over \sin(\alpha)}
\bigg(
(\cos(\alpha)+3)\delta_{ij}
-U_{ij} \sum_{\mu=1}^4{\delta_{i,j+e_\mu}
+\delta_{i,j-e_\mu}\over 2}
\bigg)
\end{equation}
Note the analogy with spatial Wilson fermions \cite{Wilson:1975id}
augmented with an {$i\gamma_4$} insertion in the Wilson term.

\section{Perturbation theory and counter-terms}
\label{perturbative}
Recent perturbative calculations \cite{Capitani:2010nn} have shown
that interactions can shift the relative positions of the poles along
the direction between them.  In other words, the parameter $\alpha$
receives an additive renormalization.  Furthermore, the form of the
action treats the fourth direction differently than the spatial
coordinates, thus breaking hyper-cubic symmetry along this direction.
This suggests three potential new counter-terms for the
renormalization of the theory.  First there is a possible
renormalization of the on-site contribution to the action proportional
to $i\overline\psi\gamma_4\psi$.  This provides a handle on the shift
of the parameter $\alpha$.  Secondly, the breaking of the hyper-cubic
symmetry indicates one may need to adjust the fermion ``speed of
light.''  This involves a combination of the above on-site term and
the strength of temporal hopping proportional to ${\delta_{i,j+e_4}
  +\delta_{i,j-e_4}}$.  Finally, the breaking of hyper-cubic symmetry
can feed back into the gluonic sector, suggesting a possible
counter-term of form $F_{4\mu}F_{4\mu}$ to maintain the gluon ``speed
of light.''  In lattice language, this corresponds to adjusting the
strength of time-like plaquettes relative to space-like ones.

Of these counter-terms, $i\overline\psi\gamma_4\psi$ is of dimension 3
and is probably the most essential.  Quantum corrections induce the
dimension 4 terms, suggesting they may be small and could partially be
absorbed into accepting a lattice asymmetry.  How difficult these
counter-terms are to control awaits simulations.

Note that all other dimension 3 counter-terms are forbidden by basic
symmetries.  For example, chiral symmetry forbids $\overline\psi\psi$
and $i\overline\psi\gamma_5\psi$ terms, and spatial cubic symmetry
removes $\overline\psi\gamma_i\psi$,
$\overline\psi\gamma_i\gamma_5\psi$, and
$\overline\psi\sigma_{ij}\psi$ terms.  Finally, commutation with
$\gamma_4$ plus space inversion eliminates $\overline\psi
\gamma_4\gamma_5\psi$.

\section{Point splitting}
\label{splitting}
The fundamental field $\psi$ can create either of the two species.
For a quantity that creates only one of them, it is natural to combine
fields on nearby sites in such a way as to cancel the other.  In other
words, one can point split the fields to separate the poles which
occur at distinct ``bare momenta.''  For the free theory, one
construction that accomplishes this is to consider
\begin{align} 
&u(q)={1\over 2}\left(1+{\sin(q_4+\alpha)\over \sin(\alpha)}
\right)\psi(q+\alpha e_4)\cr
&d(q)={1\over 2}\ {\Gamma}\left(1-{\sin(q_4-\alpha)\over \sin(\alpha)}
\right)\psi(q-\alpha e_4)\cr
\end{align}
where $\Gamma=i\gamma_4\gamma_5$ for the Karsten/Wilczek formulation.
Here I have inserted factors containing zeros cancelling the undesired
pole.  This construction is not unique, and specific details will
depend on the particular minimally-doubled action in use.  The factor
of $\Gamma$ inserted in the $d$ quark field accounts for the fact that
the two species use different gamma matrices.  This is required since
the chiral symmetry is flavored, corresponding to an effective minus
sign in $\gamma_5$ for one of the species.

It is now straightforward to transform this to position space
\begin{align}
&u_x
=
{1\over 2}{e^{i\alpha x_4}}\left(\psi_x+i\ 
{U_{x,x-e_4}\psi_{x-e_4}
-U_{x,x+e_4}\psi_{x+e_4}\over 2 \sin(\alpha)}\right)\cr
&d_x
=
{1\over 2}{\Gamma e^{-i\alpha x_4}}\left(\psi_x-i\ {
U_{x,x-e_4}\psi_{x-e_4}-U_{x,x+e_4}\psi_{x+e_4}\over 2 \sin(\alpha)}\right)\cr
\end{align}
Here I also insert gauge field factors to give simple gauge
transformation properties to the point-split field.  The various
phases inserted here serve to remove the oscillations associated with
the bare fields having their poles at non-zero momentum.

Given the basic fields for the individual quarks, one can easily
construct mesonic fields, which then also involve point splitting.  To
keep the equations simpler, I now consider the case $\alpha=\pi/2$.
For example, the neutral pion field becomes
\begin{align}
\pi_0(x)
={i\over 2}(\overline u_x\gamma_5 u_x-\overline d_x \gamma_5 d_x)=
{i\over 16}\bigg(&4\overline\psi_x\gamma_5\psi_x
+\overline\psi_{x-e_4}\gamma_5\psi_{x-e_4}
+\overline\psi_{x+e_4}\gamma_5\psi_{x+e_4}\cr
&-\overline\psi_{x+e_4}UU\gamma_5\psi_{x-e_4}
-\overline\psi_{x-e_4}UU\gamma_5\psi_{x+e_4}\bigg).\cr
\end{align}
Note that this involves combinations of fields at sites separated 
by either 0 or 2 lattice spacings.  In contrast, the $\eta^\prime$ takes
the form
\begin{align}
\eta^\prime(x)
={i\over 2}(\overline u_x \gamma_5 u_x+\overline d_x \gamma_5 d_x)=
{1\over 8}\bigg(&
\overline\psi_{x-e_4}U\gamma_5\psi_x
-\overline\psi_xU\gamma_5\psi_{x-e_4}\cr
&-\overline\psi_{x+e_4}U\gamma_5\psi_x
+\overline\psi_xU\gamma_5\psi_{x+e_4}\bigg)\cr
\end{align}
where all terms connect even with odd parity sites.  In a recent
paper, Tiburzi \cite {Tiburzi:2010bm} has discussed how the anomaly,
which gives the $\eta^\prime$ a mass of order $\Lambda_{qcd}$, can be
understood in terms of the necessary point splitting.

\section{Effective Lagrangians and lattice artifacts}
\label{effective}
As with Wilson fermions, the minimally-doubled actions have lattice
artifacts that can give rise to unusual effects.  Adding corresponding
terms to effective Lagrangians can give some insight into such
\cite{Creutz:1996bg, Sharpe:1998xm}.  With Wilson fermions the
artifacts can have one of two consequences depending on the sign of a
specific term in the effective potential.  Either the chiral
transition can become first order, or alternatively it can break up
into two transitions separated by the parity broken phase predicted by
Aoki  \cite{Aoki:1983qi}.  It appears that the latter situation is
realized with standard Wilson fermions.

Similarly, at first sight there are two possibilities here as well.
These artifacts break isospin symmetry, allowing the charged pion mass
to differ from that of the neutral pion.  The simplest situation has
the charged pions heavier.  In this case as the quark mass goes to
zero the neutral pion becomes a conventional Goldstone boson
associated with the one exact chiral symmetry of the theory.
Alternatively, if the charged pions are lighter, then before one
reaches the chiral limit their mass can go to zero and the charged
mesons would condense.  As with the Aoki phase, there would then be a
region of spontaneous parity breaking.

One qualitative difference between Wilson fermions and the
minimally-doubled case is that in the latter situation the fermion
determinant is strictly positive.  While it does not directly apply
because Lorentz invariance has been broken by the lattice artifacts,
the Vafa/Witten argument \cite{Vafa:1984xg} suggests that the first
alternative is the more likely.  Lattice simulations should be able to
settle which situation is the case.

\section {Summary}
\label{conclusions}
Minimally-doubled fermion actions present the possibility of fast
simulations while maintaining one exact chiral symmetry.  They do,
however, introduce some peculiar aspects.  An explicit breaking of
hyper-cubic symmetry allows additional counter-terms to appear in the
renormalization.  While a single field creates two different species,
spreading this field over nearby sites allows isolation of specific
states and the construction of physical meson operators.  Finally,
lattice artifacts break isospin and give two of the three pseudoscalar
mesons an additional contribution to their mass.  Depending on the
sign of this mass splitting, one can either have a traditional
Goldstone pseudoscalar meson or a parity breaking Aoki-like phase.

\end{document}